\documentclass[conference]{IEEEtran}
\IEEEoverridecommandlockouts

\usepackage{cite}
\usepackage{amsmath,amssymb,amsfonts}
\usepackage{algorithmic}
\usepackage{graphicx}
\usepackage{textcomp}
\usepackage{xcolor}

\usepackage{tikz}
\newcommand\copyrighttext{
    \footnotesize \textcopyright~2021 IEEE. Personal use of this material is permitted. Permission from IEEE must be obtained for all other uses, in any current or future media, including reprinting/republishing this material for advertising or promotional purposes, creating new collective works, for resale or redistribution to servers or lists, or reuse of any copyrighted component of this work in other works. The final version of this article is available at: https://doi.org/10.1109/IWBIS53353.2021.9631853
}
\newcommand\copyrightnotice{
    \begin{tikzpicture}[remember picture, overlay]
        \node[anchor=south, yshift=10pt] at (current page.south) {
            \fbox{
                \parbox{\dimexpr1.0\textwidth-\fboxsep-\fboxrule\relax}{\copyrighttext}
            }
        };
    \end{tikzpicture}
}

\usepackage[hyphens]{url}
\usepackage{hyperref}

\def\BibTeX{{\rm B\kern-.05em{\sc i\kern-.025em b}\kern-.08em
    T\kern-.1667em\lower.7ex\hbox{E}\kern-.125emX}}
\begin{document}

\title{
    Settlement Mapping for Population Density Modelling in Disease Risk Spatial Analysis
}

\author{
    \IEEEauthorblockN{Mirza Alim Mutasodirin}
    \IEEEauthorblockA{
        \textit{Faculty of Computer Science} \\
        \textit{Universitas Indonesia}\\
        Depok, Indonesia \\
        mirza.alim01@ui.ac.id
    }
    \and
    \IEEEauthorblockN{Rafi Dwi Rizqullah}
    \IEEEauthorblockA{
        \textit{Faculty of Computer Science} \\
        \textit{Universitas Indonesia}\\
        Depok, Indonesia \\
        rafi.dwi@ui.ac.id
    }
    \and
    \IEEEauthorblockN{Andie Setiyoko}
    \IEEEauthorblockA{
        \textit{Remote Sensing Technology} \\
        \textit{and Data Center} \\
        \textit{LAPAN}\\
        Jakarta, Indonesia \\
        andie.setiyoko@lapan.go.id
    }
    \and
    \IEEEauthorblockN{Aniati Murni Arymurthy}
    \IEEEauthorblockA{
        \textit{Faculty of Computer Science} \\
        \textit{Universitas Indonesia}\\
        Depok, Indonesia \\
        aniati@cs.ui.ac.id
    }
}

\maketitle
\copyrightnotice
\begin{abstract}
In disease risk spatial analysis, many researchers especially in Indonesia are still modelling population density as the ratio of total population to administrative area extent. This model oversimplifies the problem, because it covers large uninhabited areas, while the model should focus on inhabited areas.
This study uses settlement mapping against satellite imagery to focus the model and calculate settlement area extent.
As far as our search goes, we did not find any specific studies comparing the use of settlement mapping with administrative area to model population density in computing its correlation to a disease case rate.
This study investigates the comparison of both models using data on Tuberculosis (TB) case rate in Central and East Java Indonesia.
Our study shows that using administrative area density the Spearman's $\rho$ was considered as ``Fair'' ($0.566, p<0.01$) and using settlement density was ``Moderately Strong'' ($0.673, p<0.01$). The difference is significant according to Hotelling's t test.
By this result we are encouraging researchers to use settlement mapping to improve population density modelling in disease risk spatial analysis.
Resources used by and resulting from this work are publicly available at \url{https://github.com/mirzaalimm/PopulationDensityVsDisease}.
\end{abstract}


\section{Introduction}

One of the approaches taken to analyze the risk factors of a disease is the spatial analysis approach. This is done against diseases whose spread or increase is suspected to be influenced by spatial dimension. An associated or correlated spatial factor can be used as a predictor of a disease. Among the spatial factors, population density is often investigated. Some studies investigated its correlation against Dengue fever \cite{Garjito2020PopulationDensity}, Malaria \cite{Rakotomanana2007PopulationDensity}, Measles \cite{Qin2019PopulationDensity}, Tuberculosis (TB) \cite{Sifuna2019PopulationDensity}, and Coronavirus Disease 2019 (COVID-19) \cite{Wirawan2021PopulationDensity}.

Many large-scale research in Indonesia are still modelling population density as the ratio of total population to administrative area extent. Section \ref{sec:previousWork} will describe more about this condition. The shortcoming of this approach is that the administrative area of a regency includes large areas of forest and agricultural land. People do not exist in these uninhabited areas. Therefore, it cannot represent the problem well. People exist around their buildings, inside or outside. The model needs to focus on where people commonly exist. As an extreme example, based on this work calculation, more than 80\% of Kabupaten Situbondo, East Java province, is uninhabited area. The significance of the differences in population densities produced from the old and the proposed model will be discussed later in Section \ref{sec:result} and Table \ref{table:t-test}.

In order to focus on areas where people exist, it can ignore areas other than settlements. Settlement mapping against satellite imagery can be an approach to calculate settlement area extent and density, as the area of interest. To the best of our knowledge, no specific study compares the use of settlement mapping with administrative area to model population density in computing its correlation to a disease case rate. This study is important and contributes to investigating better approaches and their significance, encouraging researchers to use better methods and modeling.

This study used data on Tuberculosis (TB) of Central and East Java Indonesia to calculate the correlation. The hypothesis of positive correlation between population density and TB disease incidence is derived from how TB is transmitted \cite{Gavin2017Tuberculosis, Nardell2016Tuberculosis}. Our study concludes that using settlement mapping with sufficient amounts of data can improve the correlation significantly. We deliberately did not use the data on Coronavirus Disease 2019 (COVID-19) in Indonesia, which is currently being extensively researched, in order to avoid a debate about the reliability of the data. However, the results of this study may also have implications for COVID-19.

Next in Section \ref{sec:previousWork}, this article will describe previous work related to Land Use and Land Cover (LULC) classification, a research field where settlement mapping is usually conducted. It will also describe previous work in medical research investigating correlation between population density and a disease. Then it will be continued with an explanation of the data used in this study, in Section \ref{sec:data}, and how this research was conducted, in Section \ref{sec:method}. At the end, the results of this study will be discussed in Section \ref{sec:result} and will be closed with conclusion and recommendation in Section \ref{sec:conclusion}.

\section{Previous Work} \label{sec:previousWork}
In recent years, there have been many studies of Land Use and Land Cover (LULC) classification. One of the segmented classes is the settlement class. Some of the research was conducted to serve as a reference for regional development planning \cite{Saing2021LULC}, to analyze the balance of natural resources \cite{Hariyanto2020LULC}, to analyze the impact of land use change on rivers \cite{Saputra2020LULC}, to analyze the expansion of potential forest area \cite{Nurda2020LULC}, to detect illegal settlements \cite{Maula2019LULC}, to analyze settlement development \cite{Patel2015LULC}, and to predict future land change \cite{Rimba2020LULC}. Some of them used multitemporal settlement mapping data. Light Detection and Ranging (LiDAR) data are also useful for examining LULC \cite{Widyaningrum2020Lidar}.

In medical research, we still find the latest medical studies that use population density data from the Indonesian Central Statistics Agency (BPS) and the Indonesian Ministry of Health. These data are obtained from the number of population divided by administrative area extent. Using these data, Garjito et al. \cite{Garjito2020PopulationDensity} were looking for a correlation between population density and the number of dengue cases. Even the latest research on Coronavirus Disease 2019 (COVID-19) still uses these data from the BPS or the Ministry of Health. Wirawan et al. \cite{Wirawan2021PopulationDensity} and Azizah et al. \cite{Azizah2021PopulationDensity} looked for the correlation of population density with the number of COVID-19 cases. This is the condition of large-scale medical research in Indonesia lately. An example of small-scale study that used settlement mapping is a study of Astuti et al. \cite{Astuti2019RetailerDensity} that mapped buildings to calculate the density of the number of cigarette sellers in villages of Denpasar Bali Indonesia. By the conclusion of our study, the expectation is LULC researchers give more attention to supporting large-scale medical research.

\section{Data} \label{sec:data}
This study analyzes the correlation using data on 58 regencies and 15 municipalities of Central and East Java province. There are various data needed in this study. This section describes each kind of the data needed, their criteria, and their sources.

\subsection{Number of Population}
We took population data from the Indonesian Central Statistics Agency (BPS) \cite{BPS2020PopulationJateng, BPS2020PopulationJatim}. The most recent data available for municipalities and regencies in Central Java province were data for 2016, 2017, and 2020. As for East Java, data for 2018, 2019, and 2020. Therefore, this study used data on population numbers in 2020 to be equal. The weakness of this study is that we cannot choose the data year more freely.

\subsection{Administrative Area Extent}
Data on the extent of administrative areas were obtained from the BPS website for each province \cite{BPS2020AreaJateng, BPS2020AreaJatim}. The data are in units of km\textsuperscript{2}. Central Java provides data for 2018-2020. East Java provides data of 2017. Since these are data on the extent of an administrative area that do not change, the year difference is not a problem because they will always be the same from year to year.

\subsection{Tuberculosis (TB) Case}
We got the data on the number of new cases of TB disease from BPS \cite{BPS2020DiseaseJateng, BPS2020DiseaseJatim}. We had difficulty finding publicly available data, except from the website of the BPS branch of each province, and even then, not all provinces published it. As a result, we only used data from 2 provinces on the island of Java. The years of available data were 1 year different, Central Java province with data of 2019 and East Java with data of 2018. They were different from the year of population data (2020). This is a further weakness in this study.

Regarding the year difference in the data on the number of diseases in Central Java and East Java, we could not immediately combine the data from the two provinces without conducting an analysis of each province. Thus, we did 3 analyzes, namely the data for the province of Central Java only, East Java only, and the combination of both. We did these 3 analyzes to strengthen the conclusion. The data are only 1 year different, and not the data of accumulation but new cases.

Central Java BPS did not provide raw values for the number of TB cases, but rather the number of cases per 100K population. Meanwhile, East Java BPS provided raw values for the number of cases. Because the analysis process does not use raw values but ratios, the raw values of the number of TB cases in East Java must be converted to a ratio per 100K population to be equal as Central Java.

\subsection{Settlement Map}
One way to estimate the extent and the density of settlement area in a municipality and regency is to use satellite imagery. The method will be explained in more detail later in Section \ref{sec:method}. In this study, we did not do our own mapping of residential areas on satellite imagery. Mapping high-resolution satellite images with a very wide scope, 2 provinces with 58 regencies and 15 municipalities, requires a huge effort of its own which cannot be covered in this study. This study only focuses on the population density modelling experiment.

To conduct this study, we used a world-scale settlement mapping \cite{EC2020Segmentation}, the product of work by Corbane et al. \cite{Corbane2021CNN}, a global scale settlement mapping project using the proposed CNN model named GHS-S2Net \cite{Corbane2021CNN} (Deep Network for Global Human Settlement from Sentinel-2). The mapping used image data of Corbane et al. \cite{Corbane2020CompositeImage}, a combination of cloud-free images based on Sentinel-2 data for 2017-2018 available on Google Earth Engine. The CNN model was trained with 4 auto-generated datasets as described by Corbane \cite{Corbane2021CNN}. All of the training data have their own weaknesses that affect accuracy. The weakness that needs to be underlined regarding the mapping is that the training dataset came from auto-generated data.

\subsection{Basemap}
The function of the basemap here is to help see the position of the area and calculate the number of pixels of the administrative area. We used satellite imagery from Google Maps as a basemap. Google Maps provides satellite imagery with various levels of resolution, from low to very high resolution. Another reason is because of its ease of use in this study.

\subsection{Administrative Boundary}
We obtained data on the administrative boundaries of municipalities and regencies from the Indonesian Geospatial Information Agency (BIG) \cite{BIG2020RBI}. These data are in the form of shapefile format, that are geospatial vector data, and are human-annotated. These data are part of a data collection called \textit{Rupabumi Indonesia} (RBI). We took the 1:25K scale data.

\section{Method} \label{sec:method}
There are 4 steps in processing the data. The objectives are obtaining settlement density of each regency and municipality, and then computing the correlation coefficient. This section explains each of the steps in order.

\subsection{Applying the Administrative Boundary on the Basemap}
The first step to do in the process of obtaining the settlement density is to apply the administrative boundaries of the municipalities and regencies on the area of satellite imagery. We used QGIS to process the data. QGIS is a free open source geographic information system.

\begin{figure}[htbp]
\centerline{\includegraphics[width=0.3\textwidth]{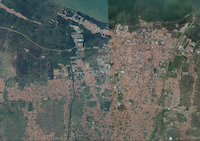}}
\caption{Basemap Layer}
\label{fig:basemapLayer}
\end{figure}

\begin{figure}[htbp]
\centerline{\includegraphics[width=0.3\textwidth]{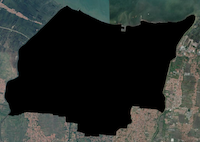}}
\caption{Administrative Boundary Layer}
\label{fig:boundaryLayer}
\end{figure}

\begin{figure}[htbp]
\centerline{\includegraphics[width=0.3\textwidth]{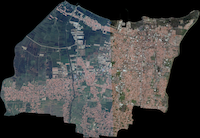}}
\caption{Inverted Administrative Boundary}
\label{fig:invertedBoundary}
\end{figure}

In QGIS we loaded satellite imagery of Google Maps on an XYZ Tiles layer. Then we loaded the shapefile of boundary data as a vector layer. As illustrated by Figures \ref{fig:basemapLayer}, \ref{fig:boundaryLayer}, and \ref{fig:invertedBoundary}, we inverted the administrative boundary for each municipality and regency. We exported each of them into a bitmap image with a resolution of 1500 dpi and a scale of 1:300K. The higher the resolution, the smaller the meters-per-pixel value and the smaller the error. We did not use higher resolution than this setting because of our memory limitation.

\subsection{Applying the Administrative Boundary on the Settlement Map}
\begin{figure}[htbp]
\centerline{\includegraphics[width=0.3\textwidth]{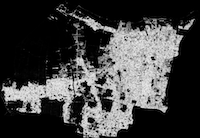}}
\caption{Settlement Map}
\label{fig:segmentation}
\end{figure}

At this step the basemap is replaced with the settlement map. The settlement map comes as a Tagged Image File Format (TIFF). We loaded the TIFF file into QGIS as a raster layer. The result of each municipality and regency obtained at this step is as illustrated in Figure \ref{fig:segmentation}. We exported each of them into a bitmap image with a resolution of 1500 dpi and a scale of 1:300K.

\subsection{Computing Population Density}
Population density value comes from the number of population divided by the area of interest (AOI). To get the area extent, we count the number of pixels in the bounded area that was generated in the first and second step. Armed with the number of pixels, population density can be computed in units of population per pixel, formulated as $d=p/x$, where $d$ is the density per pixel, $p$ is the number of population, and $x$ is the number of pixels in the AOI. Then, correlation coefficient can be computed by the data on TB case rate and the density per pixel.

The unit of population per pixel can be converted to population per square kilometers. The conversion needs the value of meters per pixel. This value is obtained by dividing the administrative area extent by the number of pixels. Supposedly, the values of meters per pixel for all municipalities and regencies are the same. However, in reality there are errors or differences. One of the most likely contributing factors is the discrepancy between the annotation of the boundary and the calculation of its extent, because it was done by different people. To deal with this problem, we take the average value without involving outliers.

However, meters per pixel is a unit of area. This is actually square meters per square pixel-side. One pixel is the square of one pixel-side. Instead of taking the average of squared values, to reduce the error, we preferred to convert them to meters per pixel-side, that is the square root of meters per pixel, as in Equation \eqref{eq:meterPerPixelside},
\begin{equation}
m=\sqrt{\frac{A\times1000^2}{x}}
\label{eq:meterPerPixelside}
\end{equation}
where $m$ is the meters per pixel-side, $A$ is the administrative area extent in km\textsuperscript{2}, and $x$ is the number of pixels. To detect outliers, we used Tukey's fence method formulated as in Equation \eqref{eq:tukeyLower} and \eqref{eq:tukeyUpper},
\begin{equation}
L=Q1-1.5(Q3-Q1)
\label{eq:tukeyLower}
\end{equation}
\begin{equation}
U=Q3+1.5(Q3-Q1)
\label{eq:tukeyUpper}
\end{equation}
where $L$ is the lower fence, $U$ is the upper fence, and $Q$ is the quartile.

We computed the settlement area extent value by squaring the average of meters per pixel-side and then multiplying it by the number of pixels of the settlement of a municipality or regency, then dividing by 10\textsuperscript{6}, as in Equation \eqref{eq:settlementArea},
\begin{equation}
\hat{A}=\frac{\hat{x}\times\bar{m}^2}{1000^2}
\label{eq:settlementArea}
\end{equation}
where $\hat{A}$ is the settlement area extent in km\textsuperscript{2}, $\hat{x}$ is the number of pixels of the settlement area, and $\bar{m}$ is the average of meters per pixel-side. To get the settlement density, the number of population is divided by the settlement area, formulated as $\hat{d}=p/\hat{A}$ where $\hat{d}$ is the settlement density, $p$ is the number of population, and $\hat{A}$ is the settlement area extent.

\subsection{Computing Correlation}
At this step, population density based on administrative area (called as old model) and settlement area (called as new model) and TB case rate data are already available. Then, the correlation between population density and the incidence of TB can be calculated. We calculated the correlation using the Spearman's rank correlation coefficient ($\rho$). The Spearman's $\rho$ is formulated as in Equation \eqref{eq:spearman},
\begin{equation}
\rho=1-\frac{6\sum_{i=1}^{n}(r_i-s_i)^2}{n^3-n}
\label{eq:spearman}
\end{equation}
where $r$ is the rank of population density, $s$ is the rank of TB case rate, and $n$ is the number of pairs of data.

We did not use Pearson's correlation coefficient because it is not suitable for data that are not normally distributed \cite{Mukaka2012Correlation, Akoglu2018Correlation}. It is sensitive to extreme values. If one has data that are not normally distributed then he should use Spearman's $\rho$ \cite{Mukaka2012Correlation, Akoglu2018Correlation}. To determine the normality of our data distribution, we used the Shapiro-Wilk normality test in the Python programming language using the shapiro function of scipy library. Many researchers consider the Shapiro-Wilk test as the best, especially for small data under 50 samples \cite{Mishra2019NormalityTest, Ghasemi2012NormalityTest}.

\begin{table*}[htbp]
\caption{Shapiro-Wilk Normality Test Results on Population Density Data\\with $n$ numbers of data and $p$-$value$ that retains or rejects null hypothesis.}
\begin{center}
\begin{tabular}{lllclc}
\textbf{Province} & \textbf{Model}& \textbf{Data} & \textbf{$n$} & $p$-$value$ & \textbf{Normality} \\
\hline
Central Java & Old Model & Regencies & 29 & $p<0.05$ & Not Normal\\
Central Java & New Model & Regencies & 29 & $p=0.134$ & Normal \\
East Java & Old Model & Regencies & 29 & $p<0.05$ & Not Normal \\
East Java & New Model & Regencies & 29 & $p<0.05$ & Not Normal \\
Central+East Java & Old Model & Regencies & 58 & $p<0.05$ & Not Normal \\
Central+East Java & New Model & Regencies & 58 & $p<0.05$ & Not Normal \\
Central+East Java & Old Model & Regencies+Municipalities & 73 & $p<0.05$ & Not Normal \\
Central+East Java & New Model & Regencies+Municipalities & 73 & $p<0.05$ & Not Normal \\
\hline
\end{tabular}
\label{table:shapiro}
\end{center}
\end{table*}

\begin{figure}[htbp]
\centerline{\includegraphics[width=0.5\textwidth]{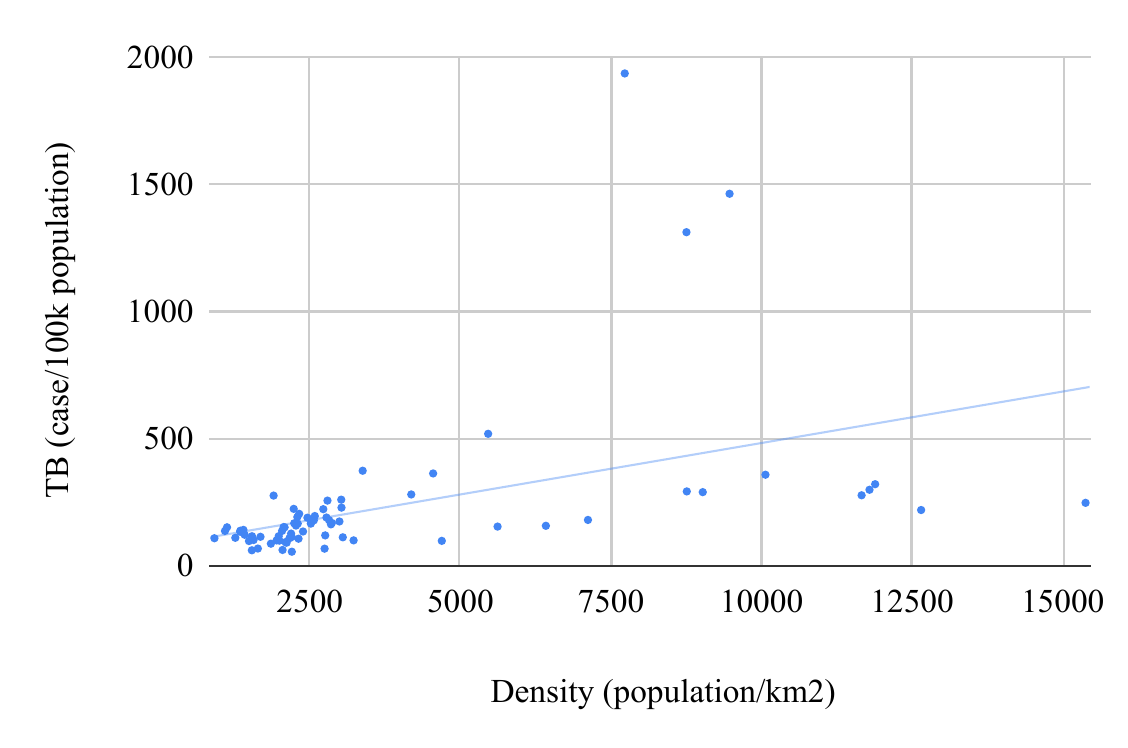}}
\caption{Data Plot of Central and East Java in New Model. It describes how population density number affects TB case rate.}
\label{fig:plot}
\end{figure}

\begin{figure}[htbp]
\centerline{\includegraphics[width=0.5\textwidth]{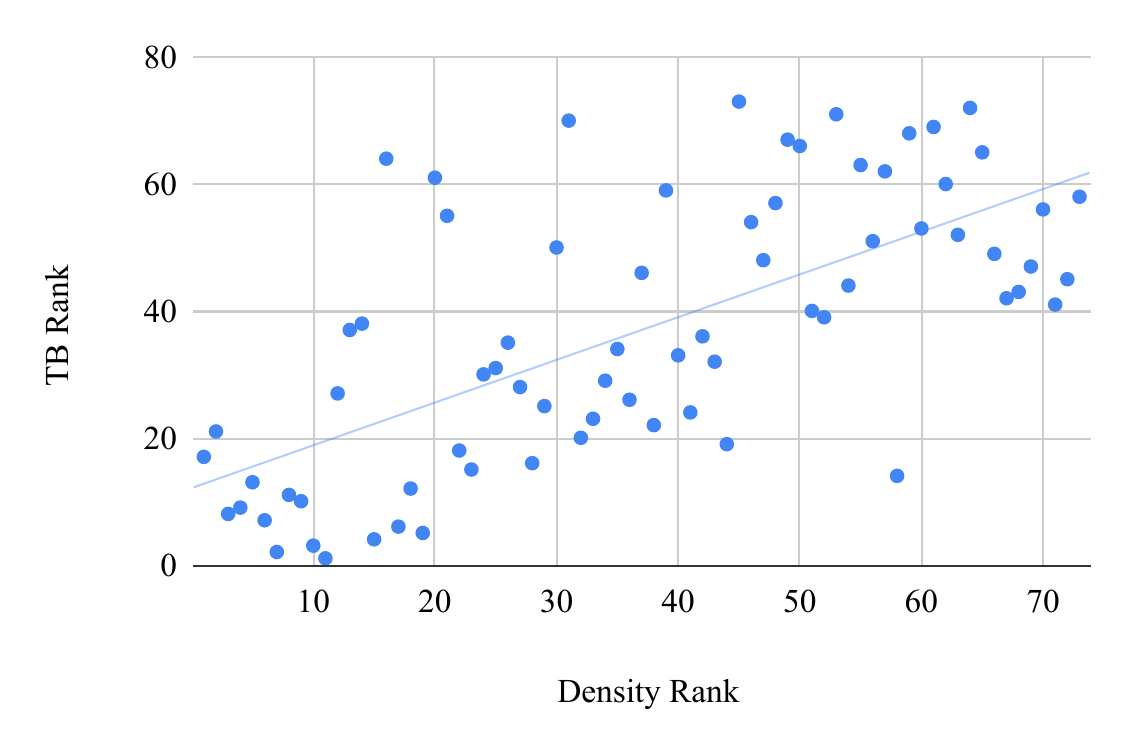}}
\caption{Rank Plot of Central and East Java in New Model. It describes how the increase in population density affects the increase in TB case rate, regardless of the raw values.}
\label{fig:rankPlot}
\end{figure}

We tested 8 variations of data as depicted in Table \ref{table:shapiro}. Null hypothesis in the Shapiro-Wilk test states that the sample data are normally distributed. We used alpha level 0.05. P-value under 0.05 means it rejects the null hypothesis. Table \ref{table:shapiro} concludes that population density data distribution tends to be not normal, especially when municipalities data are involved. Municipalities data are some of the extreme values in this dataset. Then, Figures \ref{fig:plot} and \ref{fig:rankPlot} show the difference of data plotting between raw data used in Pearson's correlation calculation and rank data used in Spearman's correlation calculation.

\begin{table*}[htbp]
\caption{CoCor Results on The Differences in Correlation Coefficients\\resulting from the old model ($\rho$) and the new model ($\hat{\rho}$) with $n$ numbers of data pairs\\and $\ddot{\rho}$ as a helper used by cocor package for assessment.}
\begin{center}
\begin{tabular}{llccccl}
\textbf{Province} & \textbf{Data} & \textbf{$n$} & \textbf{$\rho$} & \textbf{$\hat{\rho}$} & \textbf{$\ddot{\rho}$} & \textbf{Significance} \\
\hline
Central Java & Regencies & 29 & 0.254 & 0.452 & 0.490 & Insignificant \\
Central Java & Regencies+Municipalities & 35 & 0.495 & 0.608 & 0.709 & Insignificant \\
East Java & Regencies & 29 & 0.008 & 0.296 & 0.504 & Insignificant \\
East Java & Regencies+Municipalities & 38 & 0.477 & 0.610 & 0.777 & Insignificant \\
Central+East Java & Regencies & 58 & 0.329 & 0.538 & 0.628 & Significant \\
Central+East Java & Regencies+Municipalities & 73 & 0.566 & 0.673 & 0.809 & Significant \\
\hline
\end{tabular}
\label{table:result}
\end{center}
\end{table*}

\begin{table}[htbp]
\caption{T-Test Results on The Differences in Population Density Data\\from the old and the new model with $n$ numbers of data pairs\\and $p$-$value$ that retains or rejects null hypothesis.}
\begin{center}
\begin{tabular}{lllclc}
\textbf{Province} & \textbf{Data} & \textbf{$n$} & $p$-$value$ & \textbf{Significance} \\
\hline
Central Java & Municipalities & 6 & $p<0.01$ & Significant\\
Central Java & Regencies & 29 & $p<0.01$ & Significant\\
East Java & Municipalities & 9 & $p<0.01$ & Significant \\
East Java & Regencies & 29 & $p<0.01$ & Significant \\
\hline
\end{tabular}
\label{table:t-test}
\end{center}
\end{table}

\section{Result and Discussion} \label{sec:result}
Table \ref{table:result} and Table \ref{table:t-test} present the results of this study. Using the paired two-sample t-test, we measured the significance of the differences in population density resulting from the old model and the new model. The null hypothesis is that the two samples are equal. As depicted by Table \ref{table:t-test} that with a very small number of municipality data, the difference is significant. All the two-tailed p-values are under 0.01 that reject null hypothesis.

In Table \ref{table:result}, we measured correlations with 6 data variations as depicted in each row in the table. This is done to confirm the conclusion of the analysis. Column $n$ means the number of pairs of data. Column $\rho$ means Spearman's correlation coefficient between disease case rate and population density in the old model. Column $\hat{\rho}$ means Spearman's correlation coefficient between disease case rate and population density in the new model. From the measurement results of $\rho$ and $\hat{\rho}$ for each data variation, it shows that all the correlation values are higher in the new model.

We conducted a significance test on the improvement. The test compares two correlations resulting from the old model and the new model. We used the website \url{http://comparingcorrelations.org}, a graphical user interface to use cocor \cite{Diedenhofen2015Cocor} package of R programming language, to perform the test. The case in this study is two correlations based on overlapping two dependent groups \cite{Diedenhofen2015Cocor}. In this study, dependent groups mean the data are from the same provinces, while overlapping means that one of the variables is the same, that is TB case rate. In using cocor package, the correlation of the old model is symbolized as a parameter named $r.jk$ and of the new model is $r.jh$, where $r$ means the correlation coefficient, $j$ means the overlapping variable that is TB case rate, $k$ means the density in old model, and $h$ means the density in new model. Thus, $r.jk$ uses the value of $\rho$ and $r.jh$ uses the value of $\hat{\rho}$.

Column $\ddot{\rho}$ in Table \ref{table:result} means Spearman's correlation coefficient between population density in the old model and the new model. Its values are helpers needed by cocor package to assess the significance of the improvement. This correlation is symbolized as a parameter $r.kh$, where $r$ means the correlation coefficient, $k$ means the density in the old model, and $h$ means the density in the new model. The alpha level used in this test is 0.05, the confidence level is 0.95, and the null value is 0. Null value is a hypothesis of the difference between the compared correlations. Null value 0 means the two correlations are equal. This hypothesis is used as the null hypothesis in this test. We performed the one-tailed test with the alternative hypothesis that the $r.jk$ is less than $r.jh$. The tool uses various methods of significance test, one of which is Hotelling's t test, and all the results tell the same. We summarize the results in Table \ref{table:result}.

Table \ref{table:result} shows that some of them are insignificant, the test retains the null hypothesis. In our opinion, the low number of data causes this insignificance. The results of the last two rows with the number of data pairs 58 and 73, combining Central and East Java, show the significance. These results indicate that the new model is closer than the old model to the hypothesis that population density has a positive correlation against the increase in the number of TB cases. In addition, this modelling is more in line with the common sense that the analysis should focus on areas where humans generally live.

\begin{table}[htbp]
\caption{Correlation Coefficient Value Interpretation}
\begin{center}
\begin{tabular}{cc}
\textbf{Coefficient Value} & \textbf{Strength} \\
\hline
$\rho \geq 0.8$ & Very Strong \\
$0.6 \geq \rho < 0.8$ & Moderately Strong \\
$0.3 \geq \rho < 0.6$ & Fair \\
$\rho < 0.3$ & Weak \\
\hline
\end{tabular}
\label{table:interpret}
\end{center}
\end{table}

In interpreting the correlation coefficient in medical research, we refer to the interpretation of Y. H. Chan \cite{Chan2003Correlation, Akoglu2018Correlation}. ``Very Strong'' correlation is at the level of 0.8 and above. A value between 0.6 and 0.8 means ``Moderately Strong''. While values between 0.3 to 0.6 are ``Fair'' and below 0.3 are ``Weak''. Nevertheless, correlation is not a causal relationship. High value of correlation coefficient cannot be interpreted as a strong causal relationship. Causation requires separate analysis which will be more difficult \cite{Hung2017Correlation}. However, the correlation can be another clue to lead to an analysis of the cause of the high number of TB incidences. In an emergency, correlation can be an alternative to make predictions by looking at the trendline.

\section{Conclusion and Recommendation} \label{sec:conclusion}
This study contributes to investigating a better approach and modelling. It concludes that using settlement maps and settlement density with sufficient numbers of data has an indication to be more representative to model population density in the context of disease risk spatial analysis. The difference compared to the old model is statistically significant. Another conclusion this study can draw is that in the provinces of Central and East Java the population density has a "Moderately Strong" positive correlation against the increase in the number of TB cases, with Spearman's $\rho$ value of 0.673 ($p<0.01$). By the conclusion of this study, we recommend holding a large-scale settlement mapping on the territory of Indonesia periodically. Thus, medical researchers can conduct research in a more representative modelling. Multitemporal settlement maps also have their own benefits \cite{Patel2015LULC}.

Future work that can be held is to build a human-annotated settlement map dataset from satellite imagery specifically for Indonesian territory. This is done to overcome the weakness of the auto-generated training data used by Corbane \cite{Corbane2021CNN}. The dataset is suggested to be the representative of various criteria of settlements and nonsettlements in each province and culture. In addition, other studies with the same method as this study can be carried out by improving the weaknesses of the data, including increasing the amount of data to be much larger, or applying it to different research fields. Advanced capabilities such as building classification can be added to distinguish between types of buildings, which ones are densely populated and which ones are sparsely populated. In this way, researchers may be able to give different weights to each type of building.

\bibliographystyle{./IEEEtran}
\bibliography{./conference_041818}

\end{document}